\documentstyle[prd,aps,twocolumn,psfig]{revtex}

\draft
\begin{document}

\twocolumn[
\hsize\textwidth\columnwidth\hsize


\begin{center}
{\large\bf{VHE Neutrinos and Gamma-Rays as Probes for Anisotropy
of\\ Arrival Directions of EeV Cosmic Rays}}
~\\
~\\
Shigehiro Nagataki and Keitaro Takahashi\\
~\\
   Department of Physics, School of Science,
    the University of Tokyo,\\
    Hongo 7-3-1, Bunkyo-ku, Tokyo 113-0033, Japan
\end{center}

\begin{abstract}
We estimate the fluxes of very high energy (VHE) gamma-rays and
neutrinos from the Galactic Center
which are by-products of the hadronic interactions
between primary protons which are accelerated up to $\sim 10^{19}$eV
and ambient matter. We find that these signals
depend strongly on the spectrum index of the primary protons.
This means that detection of these signals from the Galactic Center
will constrain the spectrum of the primary particles, which, in turn,
should constrain the acceleration mechanism at the Galactic Center.
We show that GeV-TeV gamma-rays will be detected by the ground-based
Cherenkov telescopes such as CANGAROOIII and the next-generation
satellites such as GLAST. 
Thus the acceleration region of the CRs below $\sim \; 10^{19}$eV will
be determined accurately by gamma-ray telescopes,
which should help our understanding on the particle acceleration.
We estimate the event rate of VHE neutrinos
for the planned 1 km$^{2}$ detectors such as IceCube, ANTARES and NESTOR.
The event rate also depends on the spectrum index of the primary protons.
In the optimistic case, the event rate amounts to $\sim 10^4$ events
per year and the signals from the Galactic Center dominate the atmospheric
neutrino background.
\end{abstract}
\vskip2pc]

\vskip 1cm

\section{Introduction}

The question of origin of cosmic rays (CRs) continues to be 
an unsolved problem even after almost ninety years of research
since the announcement of their discovery in 1912. Although the
general aspects of the question of CR origin are regarded as fairly
well-understood now, major gaps and uncertainties still remain,
the level of uncertainty being in general a function that
increases with energy of the CRs.

It is generally accepted that the majority of CRs observed on the
Earth are accelerated within our galaxy. Various mechanisms have
been proposed for their acceleration, including the very efficient
shock acceleration in supernova explosions. On the other hand, almost
all proposed galactic mechanisms have difficulty in explaining the highest
energy CRs ($\sim 10^{20}$eV), and the lack of any likely candidates in our
galaxy suggests the existence of the extragalactic sources for the ultra-high
energy CRs. Thus there should be a transition at some energy where the Galactic
components decreases and extragalactic components become to dominate.

Recently, a high statistics search for anisotropy has been reported
by the Akeno Giant Air Shower Array (AGASA) collaboration \cite{AGASA}.
They have found significant anisotropy of $\sim 4 \%$ around
$10^{18}$eV.
A sky map of the CR density showed that this anisotropy was apparently
caused by an excess of 0.1 sr (diameter $\sim$ 1.4 kpc at 8 kpc)
in a direction close to the Galactic Center,
with a smaller excess from the direction of the Cygnus region of
the Galactic plane. The excess was a $4.5 \sigma$ and $3.9 \sigma$,
respectively, above the expected isotropic flux.
Data from the SUGAR detector also confirmed the existence of an
excess of $10^{18}$eV CRs from a direction close to the Galactic
Center \cite{SUGAR}. They found that the signal is consistent with
that from a point source, although the point source seems to be not at
the Galactic Center itself.
Some evidence for a galactic plane excess was also seen in data
from the Fly's Eye experiment over a similar energy range~\cite{Fly'sEye}.
This was a broad-scale study, and no attempt was made to identify
whether any particular Galactic longitudes were responsible
for the excess.

We consider that the special candidate for the accelerator around the
Galactic Center will be the super massive black hole at the Galactic
Center. However, there is a problem. The position of the peaks of the
signals in both of AGASA and SUGAR data are close to the Galactic
Center, but not at the Galactic Center. How should we interpret this fact?
There will be two explanations. One is that AGASA and SUGAR may suffer from
a systematic pointing error in local coordinates. In fact, it is reported
that the peaks of the signals in SUGAR data are displaced from the peaks
of AGASA data by about $6^{\circ}$, which is comparable with the displacement
between the peaks of SUGAR data and the Galactic Center. Another explanation
is that there may be a compact system near the Galactic Center, such as a
neutron star or a black hole like the Cygnus X-1~\cite{bednarek02}.
In this study, we mainly consider the possibility of the former
interpretation, because extraordinary energetic cosmic rays should be
produced in an extraordinary environment such as a region around a
massive black hole whose mass is about $10^6 M_{\odot}$, which
exists only at the Galactic Center.

Observations of the high energy CR anisotropy, when taken together with
studies of the CR spectral shape and CR composition, can yield very important
clues to the nature of the high energy CRs. 
One possible explanation of the observed anisotropy is due to
propagation of high energy protons which are much produced at the
Galactic Center.


Another possible explanation is that the anisotropy is due to
neutron primary particle. Neutrons of $10^{18}$eV have a Lorenz
factor of $10^{9}$ and their decay length is about 10kpc. Therefore
they can propagate linearly from the Galactic Center without decaying.
In this study, we investigate the possibility
that the observed excess near the Galactic Center is composed of
neutrons produced by collisions between ambient matter 
and protons. We assume that protons are accelerated up to 
$\sim 10^{19}$eV at the Galactic Center, where the inter-stellar matter 
is concentrated in a narrow layer ($R_{p} \sim$ 50pc) of molecular gas 
for a total mass of $\sim 10^{8} {\rm M}_{\odot}$.
We show that much of the primary protons interact with this ambient
matter and the amount of the neutrons becomes not to be negligible.
The possibility of detection of gamma-rays and neutrinos, which are
by-products of the above collision, is also discussed.
The detection of GeV-TeV gamma-rays and TeV-PeV neutrinos
from the accelerated region at the Galactic Center should strongly
support the solution of neutron primary particle. 
We show that the fluxes of GeV-TeV gamma-rays and TeV-PeV
neutrinos 
depend strongly on the spectrum index of the primary protons.
This means that detection of these very high energy (VHE) signals
from the Galactic Center
will constrain the spectrum of the primary particles, which, in turn,
should constrain the acceleration mechanism at the Galactic Center.

In section II, we estimate the flux ratio of high energy neutron to proton.
Gamma-ray flux, neutrino flux, and their detectability are estimated
in section III.
Discussion and conclusion are presented in section IV.

\section{Model and neutron flux}

We use CR data from AGASA \cite{AGASA} and assume that the observed
anisotropy, which is the most significant in the bin E5 ($1.0 - 2.0
\times 10^{18}$eV), is energy-independent in the energy region
the bin 4 - 7 ($0.5 - 8.0 \times 10^{18}$eV).
In the significance map with beam size of 20$^{\circ}$, a
4.5$\sigma$ excess (obs./esp. = 506/413.6) near the Galactic
Center region is reported, which corresponds to about 22$\%$ ($\equiv f$)
enhancement.
It is then likely that protons are accelerated up to $\sim 8 \times 10^{18}$
eV at the Galactic Center. These protons interact with ambient photons
and protons during propagation.

\subsection{$p\gamma$ collision}
The neutrons are produced as the by-product of photo-meson production,
$
p + \gamma \rightarrow n + \pi^{+}.
$
The total cross section for this collision, $\sigma_{p\gamma}$,
is characterized by a resonance region near the threshold,
$
E_{p} E_{\gamma} = 0.2 {\rm GeV}^{2},
$
 due to,
$
p + \gamma \rightarrow \Delta^{+} \rightarrow n + \pi^{+}.
$
Here $E_{p}$ and $E_{\gamma}$ is the proton and target photon energy,
respectively. From this formula, the target photons of $10^{18}$eV
proton is the photons of energy,
$
E_{\gamma} = 0.2 \; \left(10^{18} {\rm eV} / E_{p} \right) \; {\rm eV}.
$
The peak cross section is $\sigma_{p\gamma} \sim 250 \mu{\rm b}$
\cite{SOPHIA}.

The photon density $n_{\gamma}$ of this energy range around the Galactic
Center can be estimated from COBE observation \cite{COBE},
\begin{equation}
L_{\gamma}(0.2 {\rm eV}) \sim 3 \times 10^{44} {\rm erg} \; {\rm s}^{-1},
\end{equation}
as,
\begin{eqnarray}
n_{\gamma} & \sim & \frac{L_{\gamma} R}{(4\pi \; R^{3}/3)\times 0.2{\rm eV} 
\times c} 
\nonumber \\
           & \sim & \frac{3 \times 10^{44}{\rm erg} \; {\rm s}^{-1} \times
1.4 {\rm kpc}}{(4\pi \; (1.4 {\rm kpc})^{3}/3) \times 0.2 {\rm eV} \times 
3 \times 10^{10} {\rm cm}
\; {\rm s}^{-1}} \nonumber \\
           & \sim & 420 {\rm cm}^{-3}.
\end{eqnarray}
As is shown in the following subsection, $p\gamma$ collision
does not contribute the neutron flux compared to $pp$ collision.

\subsection{$pp$ collision}
Accelerated protons lose their energy by collision with interstellar
matter,
\begin{equation}
p + p \rightarrow \pi^{\pm} + X. \label{pp_collision}
\end{equation}
Here a fraction $\sim 0.5$ of the proton energy goes into pion
production, and the nucleon X emerging after the interaction
is a neutron $\sim 50 \%$ of the time \cite{Protheroe}.
The cross section $\sigma_{pp}$ is $\sim 50$ mb at the energy
range considered here.
Furthermore, the decay chain of charged pions generate high-energy 
neutrinos and we discuss detectability of these neutrinos later.

In the central 600pc, a sky area often referred as the Nuclear Bulge,
the interstellar matter is concentrated in a narrow layer ($R_{p} \sim$ 50pc)
of molecular gas for a total mass of $\sim 10^{8} {\rm M}_{\odot}$ 
\cite{density}.
The number density $n_{p,{\rm bulge}}$ is $\sim 10^{4} {\rm cm}^{-3}$.
On the other hand, the interstellar matter in the disk has
density $n_{p,{\rm disk}} \sim 1 {\rm cm}^{-3}$. Then the ratio of $pp$ 
collision rate
in the Nuclear Bulge $r_{\rm bulge}$ to that in the disk $r_{\rm disk}$ is,
\begin{eqnarray}
\frac{r_{\rm bulge}}{r_{\rm disk}} 
& = & \frac{n_{p,{\rm bulge}} \times R_{p}}{n_{p,{\rm disk}} \times d} 
\nonumber \\
& = & \frac{10^{4}{\rm cm}^{-3}  \times 50 {\rm pc}}
           {1 {\rm cm}^{-3}  \times 8 {\rm kpc}} \nonumber \\
& = & 62.5,
\end{eqnarray}
where $d \sim 8$kpc is the distance from the Earth to the Galactic Center.
Therefore, $pp$ collision in the Nuclear Bulge dominates.

On the other hand, the ratio of $pp$ collision rate
in the Nuclear Bulge $r_{\rm bulge}$ to that of $p\gamma$ collision rate
$r_{\gamma}$ is,
\begin{eqnarray}
\frac{r_{\rm bulge}}{r_{\gamma}} 
& = & \frac{10^{4}{\rm cm}^{-3}  \times 50 {\rm pc} \times 50 {\rm mb}}
           {420 {\rm cm}^{-3}  \times 1.4 {\rm kpc} \times 250 \mu{\rm b}} 
\nonumber \\
& \sim & 170.
\end{eqnarray}
Therefore, $pp$ collision in the Nuclear Bulge dominates.

From here, we take into account only $pp$ collision in the Nuclear Bulge
which dominates if the CRs of the energy range below $\sim 10^{19}$eV
originate mainly from the Galactic source.
The mean free path $\lambda_{p}$ of a high-energy proton is then,
\begin{equation}
\lambda_{p} \sim \frac{1}{\sigma_{pp} n_{\rm bulge}} 
            \sim 600 {\rm pc}.
\label{eq:mfp}
\end{equation}
Therefore, 1/12 of high-energy protons interact with interstellar matter
in the Nuclear Bulge, when we neglect the effect of magnetic field.
Thus we can conclude that the amount of high
energy neutrons can not be negligible (at least $\sim 5 \%$) as long as the
source is the Galactic Center itself.

We have neglected the effects of the magnetic field at the Galactic Center,
since the amplitude and the configuration of magnetic field at the
Galactic Center are poorly known observationally. This is justified if
the magnetic field is
weak enough ($B_{\rm GC} \lesssim 20 {\rm \mu G}$) or the correlation length
is short ($l_{\rm c} \lesssim 1 {\rm pc}$). 
However, if the narrow layer of molecular gas was formed by
contracting of interstellar matter, the energy density of magnetic field there
can be as large as,
\begin{equation}
\frac{B^{2}_{\rm bulge}}{8 \pi} \sim
\frac{n_{p, {\rm bulge}}}{n_{p, {\rm disk}}}
\frac{B^{2}_{\rm disk}}{8 \pi}
\sim
10^{4} \frac{B^{2}_{\rm disk}}{8 \pi}.
\end{equation}
So,
\begin{eqnarray}
B_{\rm bulge} \sim 300 {\rm \mu G}.
\end{eqnarray}
If the correlation length of the magnetic field is much shorter
than the proton Lamor radius ($l_{p} \sim 3 {\rm pc} 
(300 {\rm \mu G}/B_{\rm bulge}) (E_{p}/10^{18} {\rm eV})$),
the magnetic field does not affect the propagation of protons
and the above results do not change. Otherwise, the path length of
protons in the Nuclear Bulge substantially increase and 
it can be estimated by diffusion approximation as,
\begin{equation}
\frac{3 r_{\rm bulge}^{2}}{l_{p}} \sim 2 {\rm kpc} 
\left(\frac{B_{\rm bulge}}{300 {\rm \mu G}}\right)
\left(\frac{10^{18} {\rm eV}}{E_{p}}\right).
\end{equation}
Since this is larger than the proton mean free path (\ref{eq:mfp}),
all of the protons interact with interstellar matter in the Nuclear
Bulge. In this case, all the CRs which come from the direction of the 
Galactic Center are neutrons.

\section{Gamma-ray flux, neutrino flux, and their detectability}
\label{gamnudet}

In the previous section, it is shown that the amount of high energy
neutrons can not be negligible (at least $\sim 5 \%$) as long as the
source is the Galactic Center itself. 
In this section, we estimate the
flux of GeV-TeV gamma-rays and TeV-PeV neutrinos which are the products of
interactions between high energy cosmic rays and interstellar matter,
assuming that the enhancement of the intensity ($f$=0.22) of the 
cosmic rays at $10^{18}$eV is composed of neutrons.
We also investigate the case of $f$=1.0 as an extreme case.
This case corresponds to the case when the effects of the magnetic
fields at the Galactic Center are not negligible, which was described
at the end of the previous section.

\subsection{$\gamma$ ray flux and detectability}

Gamma-rays are produced by $pp$ collisions, which are decay products of
$\pi^{0}$, may be detected as signals from the Galactic Center.
The high energy protons lose $\sim 50 \%$ of their energies
per each interaction.
First, we assume that the number flux of protons
[cm$^{-2}$ s$^{-1}$ sr$^{-1}$ GeV$^{-1}$] at the Earth
from the Galactic Center can be written as
\begin{eqnarray}
    f(E_{p}) =  k_1 \left( \frac{1 \rm GeV}{E_p} \right)^{\gamma}.
    \label{eq:1}
\end{eqnarray}
Here we assume that protons from the Galactic Center dominate
only in the narrow range around $10^{18}$eV over those from the
other CR sources.
It is not necessary for this spectrum to obey the form of the average
energy spectrum of the cosmic rays in our galaxy. Thus, we leave
$\gamma$ as a free parameter.
Using this spectrum, we can estimate the number flux of
gamma-rays [cm$^{-2}$ s$^{-1}$ GeV$^{-1}$]
at the Earth from an emitting region whose solid angle is $d \Omega$. 
The expected number flux is~\cite{nagataki02}
\begin{eqnarray}
    f(E_{\gamma}) &=& \frac{1}{4 \pi D^2} \left. 
     \frac{dN_{\gamma}}{dE_{\gamma}dt}
    \right|_{E_{\gamma}}  = 
    \frac{ N^2(E_p)}{4 \pi D^2}\left. \frac{24dN_p}{dt d E_{p}}
    \right |_{E_p} \\ 
    &=& 24N^2(E_p)  k_1 k_2 \left( \frac{1 \rm GeV}{E_p} \right)^{\gamma}
        d \Omega,
    \label{eq:2}
\end{eqnarray}
where $N(E_p) = 0.61 + 0.56 \ln (1.88E_p) + 0.129 \ln^2(1.88 E_p)$ is
mean value of the charged-particle multiplicity as a function of
$\sqrt{s}$~\cite{breakstone84}. It is noted that the mean energy of
gamma-rays can be written as $E_p/24N(E_p)$ when the average number of
$\pi^{\circ}$, $\pi^{+}$ and $\pi^{-}$ produced per one inelastic
interaction is same with each other. The parameter $k_2$ is introduced
to represent the fraction of cosmic rays which interact with ambient matter.

On the other hand, the number flux of neutron at the Earth from the 
Galactic Center [cm$^{-2}$ s$^{-1}$ GeV$^{-1}$] can be written as
\begin{eqnarray}
    f(E_{n})
    &=& k_1 k_2 \left( \frac{1 \rm GeV}{2E_n} \right)^{\gamma}
    \exp \left\{ -0.923 \left( \frac{10^9 \rm GeV}{E_n}  \right) \right\}
    d \Omega.
    \label{eq:3}
\end{eqnarray}
Here we assumed that the probability that a neutron emerges after
one interaction is 50$\%$. In this expression, the lifetime of a neutron
at its rest frame (= 886.7s) is taken into account.

\begin{figure}[htbp]
\centerline{\psfig{figure=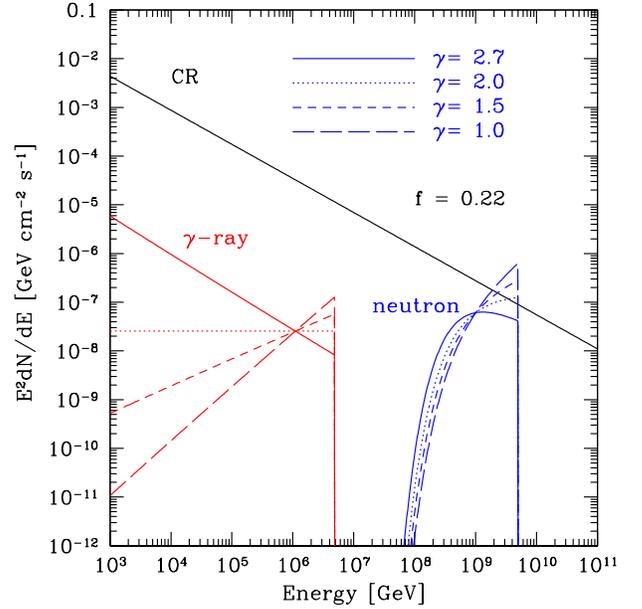,width=8.5cm}}
\caption{
Number flux of gamma-rays and neutrons
at the Earth from the Galactic Center as a function of $\gamma$.
The opening angle is set to be 20$^{\circ}$.
The maximum energy of protons at the Galactic Center is assumed to be
10$^{10}$GeV.
The enhancement of the cosmic rays at 1EeV at Galactic Center is assumed
to be composed of neutron ($f$=0.22).
Background cosmic rays is also shown for comparison.
}
\label{fig1}
\end{figure}

As for the number flux of cosmic rays
[cm$^{-2}$ s$^{-1}$ sr$^{-1}$ GeV$^{-1}$], it can be fitted in wide range as
\begin{eqnarray}
    f(E_{\rm CR})
    &=& 1.46\times \left( \frac{1 \rm GeV}{E_{\rm CR}}  \right)^{2.7}.
    \label{eq:3.2}
\end{eqnarray}
Since we consider the case the neutron flux from the Galactic Center
is 0.22 times ($\equiv f$) that of the flux of cosmic rays at $10^{18}$eV,
we can derive the relation as
\begin{eqnarray}
k_1 k_2 = 8.1 \times 10^{-1} \times 2^{\gamma} \left( 
\frac{1 \rm GeV}{10^9 \rm GeV}  \right)^{2.7 - \gamma}.
    \label{eq:4}
\end{eqnarray}
Thus, we can determine the number flux of gamma-rays and neutrons
at the Earth from the Galactic Center as a function of $\gamma$.
The results is shown in Fig.~\ref{fig1}. The opening angle is set
to be 20$^{\circ}$~\cite{bednarek02}.
In this figure, the maximum
energy of protons at the Galactic Center is assumed to be 10$^{10}$GeV.
Due to this cut-off, the number fluxes of gamma-rays and neutrons 
also suffer a steep cut off. It is clearly shown that the intensity
of gamma-rays becomes lower and that of neutrons becomes higher 
as the index $\gamma$ becomes larger. Background cosmic rays is also shown
in this figure.

\begin{figure}[htbp]
\centerline{\psfig{figure=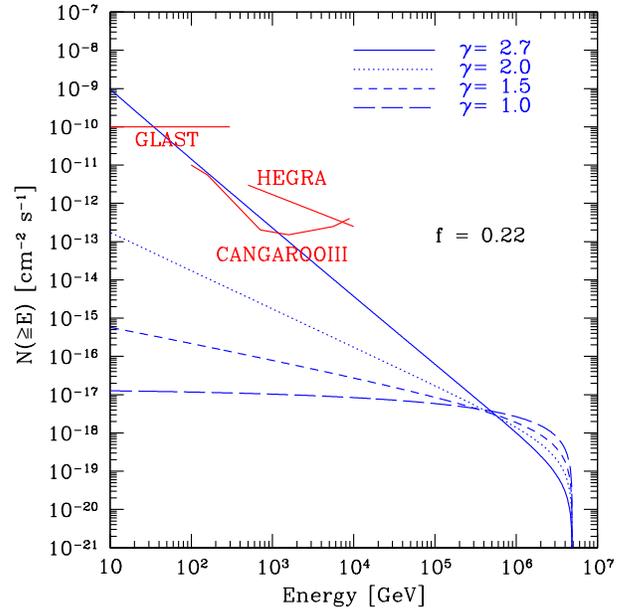,width=8.5cm}}
\caption{
The integrated spectra of gamma-rays.
The opening angle is set to be 0.1$^{\circ}$.
The enhancement of the cosmic rays at 1EeV at Galactic Center is assumed
to be composed of neutron ($f$=0.22).
The sensitivities of the ground-based Cherenkov telescopes
(HEGRA and CANGAROOIII) for a 50-hour exposure on a single source
and GLAST sensitivity for one year of all-sky survey are also
shown for comparison.
}
\label{fig2}
\end{figure}

\begin{figure}[htbp]
\centerline{\psfig{figure=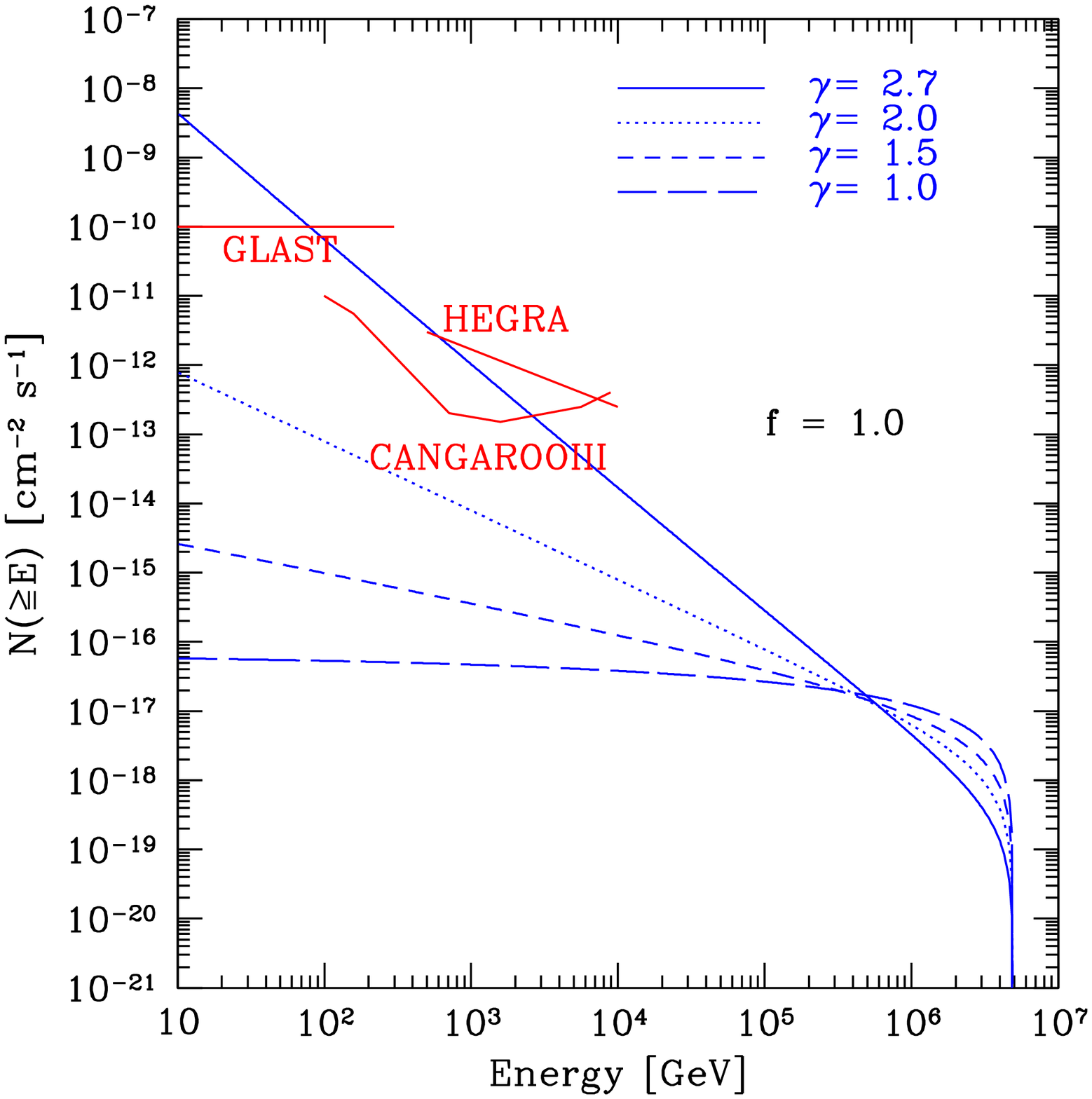,width=8.5cm}}
\caption{
Same as Fig.2 but for $f=1.0$ is adopted for the chemical
composition of the cosmic rays at Galactic Center.
}
\label{fig2b}
\end{figure}

The integrated spectra of gamma-rays are shown in Fig.~\ref{fig2}.
The opening angle is set to be 0.1$^{\circ}$.
The sensitivities of the ground-based Cherenkov telescopes
(HEGRA and CANGAROOIII) for a 50-hour exposure on a single source
and GLAST sensitivity for one year of all-sky survey are also
shown for comparison. In Fig,~\ref{fig2b}, the integrated spectra
of gamma-rays for the case that all of the cosmic rays around
10$^{18}$eV is composed of neutron is shown.
GLAST and CANGAROOIII will detect the diffuse gamma-rays around the
Galactic Center even if only the enhancement of the cosmic rays at 1EeV
is composed of neutron ($f$=0.22), as long as the spectrum of cosmic rays
at the Galactic Center is soft ($\gamma = 2.7$).
Only in the most optimistic case in which all of cosmic rays are composed
of neutrons ($f$=1.0) and the spectrum is soft ($\gamma = 2.7$), HEGRA might
be able to detect the signals from the decays of $\pi^{\circ}$'s from the
Galactic Center region, which shows that no bound is violated for
gamma-ray flux in the range of GeV-TeV energies.

%

\subsection{Neutrino flux and detectability}\label{neufludet}

High-energy neutrinos are also generated in the decay chain of the charged 
pions, which are produced in the $pp$ collision (\ref{pp_collision}),
\begin{eqnarray}
\pi^{\pm} & \rightarrow & \mu^{\pm} + \nu_{\mu}/\bar{\nu}_{\mu}, \\
\mu^{\pm} & \rightarrow & e^{\pm} + \nu_{e}/\bar{\nu}_{e} + 
\bar{\nu}_{\mu}/\nu_{\mu}.
\label{eq:5}
\end{eqnarray}

\begin{figure}[htbp]
\centerline{\psfig{figure=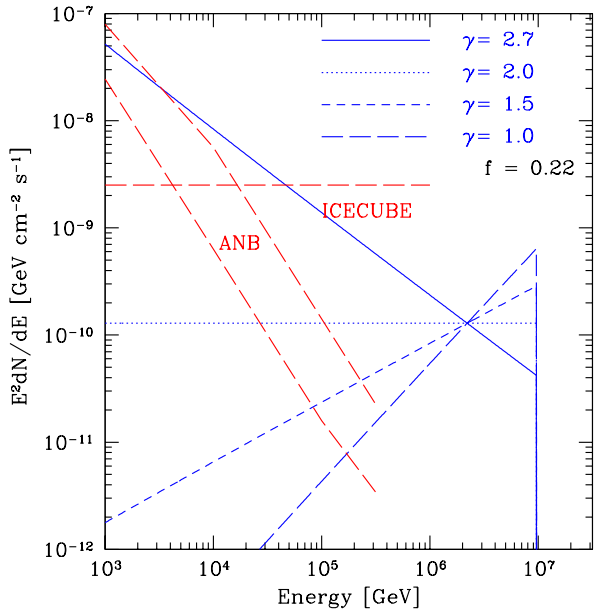,width=8.5cm}}
\caption{
The flux of $\nu_{\mu}$'s produced at the Galactic Center.
The enhancement of the cosmic rays at 1EeV at Galactic Center is assumed
to be composed of neutron ($f$=0.22).
The flux of the atmospheric neutrino background
(ANB) within a $1^{\circ}$ of the source and the three year sensitivity
of IceCube detector are also shown for comparison.
}
\label{fig3}
\end{figure}

\begin{figure}[htbp]
\centerline{\psfig{figure=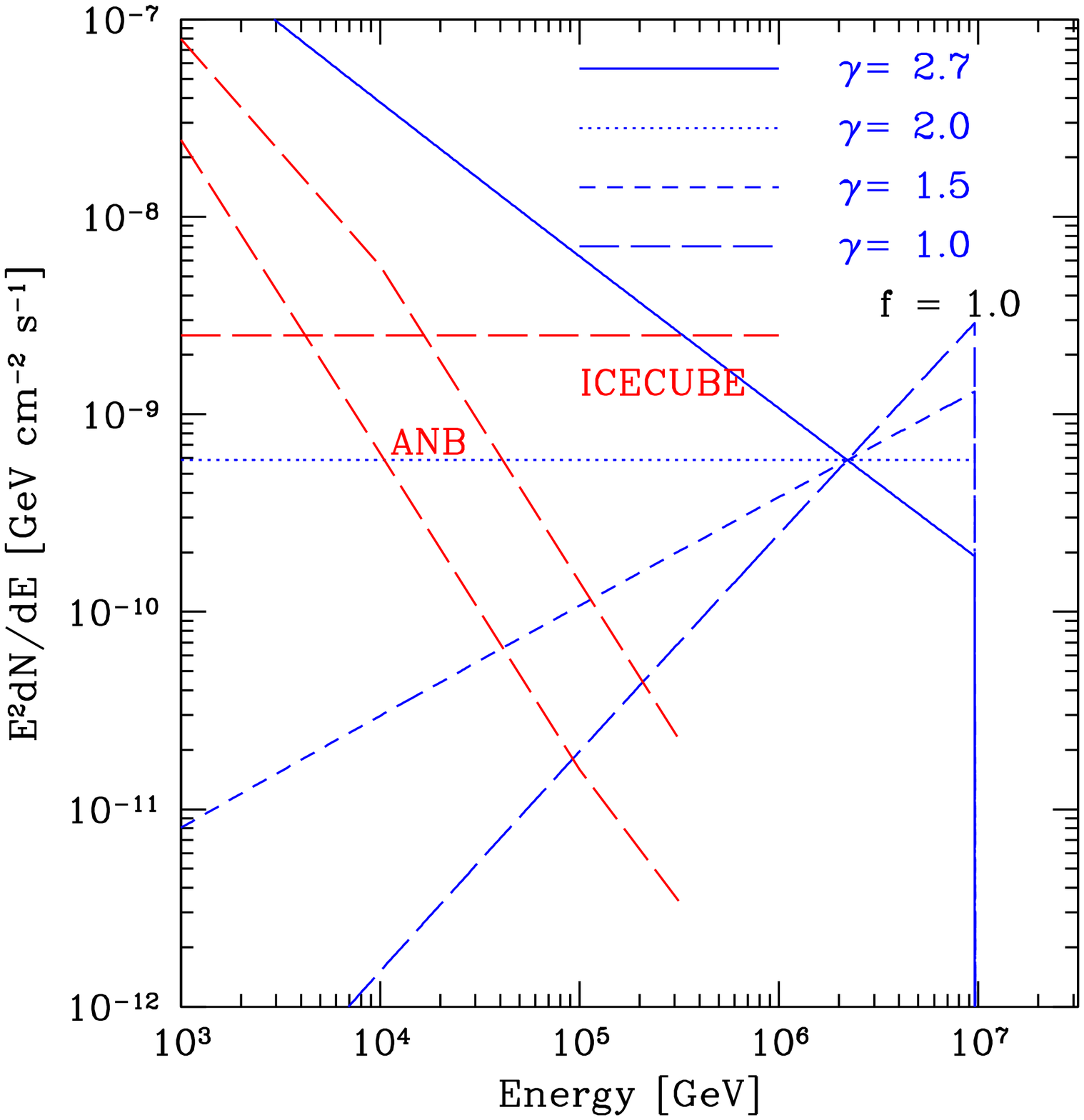,width=8.5cm}}
\caption{
Same with Fig.4, but for $f=1.0$ is adopted for the chemical
composition of the cosmic rays at Galactic Center.
}
\label{fig3b}
\end{figure}

The number flux of
$\nu_{\mu}$'s [s$^{-1}$ cm$^{-2}$ GeV$^{-1}$]
at the Earth from a single source can be written as
\begin{eqnarray}
    f(E_{\nu_{\mu}}) 
    = 12N^2(E_p)  k_1 k_2 \left( \frac{1 \rm GeV}{E_p} \right)^{\gamma}
      d \Omega,
    \label{eq:6}
\end{eqnarray}
where $E_p$ = 12$N(E_p)E_{\nu_{\mu}}$. 
The flux of $\nu_{\mu}$'s produced at the Galactic Center for the case of
$f=0.22$ is shown in Fig.~\ref{fig3}.
The flux of the atmospheric neutrino background
(ANB) within a $1^{\circ}$ of the source and the three year sensitivity
of IceCube detector are also shown for comparison.
The resulting neutrino flux can be larger than that of ANB
when the spectrum is soft ($\gamma=2.7$) in the energy range
$E_{\nu_{\mu}} \ge 3 \times 10^3$GeV. Also, the flux can be over the detection
limit of IceCube. However, if the spectrum is hard, the signals are
too difficult to be detected. This tendency is same even if 
all of the cosmic rays at the galactic center is composed
of neutrons ($f=1.0$), which is shown in Fig.~\ref{fig3b}.
In the extreme case ($f=1.0$ and $\gamma=2.7$), the flux is quite
high, although this flux is still under the limit obtained with
AMANDA-B10~\cite{hill01}.


We estimate the event rate (yr$^{-1}$)
at the planned $1 {\rm km}^{2}$ detectors of
high-energy neutrinos include IceCube, ANTARES, NESTOR \cite{NESTOR},
and NuBE \cite{NuBE}. Neutrinos are detected by observing optical Cerenkov
light emitted by neutrino-induced muons. The probability that a muon neutrino
will produce a high-energy upward moving muon with energy above 2 GeV in
a terrestrial detector is approximately given by the ratio of the muon
range to the neutrino mean free path~\cite{Gaisser}.
The result can be fitted as
\begin{eqnarray}
	P(E_{\nu}) &=& 1.3 \times 10^{-9} \left( \frac{E_{\nu}}{1 \; {\rm GeV}}
                     \right) ^{2}
 \;\; E_{\nu} \le 1 \; {\rm TeV} \\
		   &=& 1.3 \times 10^{-9} \left( \frac{E_{\nu}}{1 \; {\rm GeV}}
                     \right) \;\; E_{\nu} \ge 1 \; {\rm TeV}
    \label{eq:20}
\end{eqnarray}
where $\beta =2$ for $E_{\nu} \le 1$ TeV and $\beta =1$ for $E_{\nu} \ge 1$
TeV. 
Thus the expected event rate R [yr$^{-1}$] is (over 2$\pi$ sr)
\begin{eqnarray}
  R = 3\times10^{17}\left( \frac{d}{1 \; \rm {km}} \right)^2 
     \int_{2 \rm{GeV}}
       d E_{\nu} 2 \frac{2\pi}{4 \pi}
       f(E_{\nu}) P(E_{\nu}),
    \label{eq:21}
\end{eqnarray}
where $d$ is the diameter of the detector. The factor 2 in the integral
is introduced to count in both of contributions from $\nu_{\mu}$'s and
$\bar{\nu}_{\mu}$'s.
The results is shown in Table~\ref{tab:table0}.
As for the case of
$\gamma = 2.7$, the event rate is quite high. In this case,
the signals will be detected by the planned $1 {\rm km}^{2}$ detectors
in the near future. As for the other cases, the event rate is quite
low and the significant detection of the signals from the Galactic Center
seems to be too difficult.

\section{Discussion and conclusion}

In this study, we show that the fluxes of GeV-TeV gamma-rays and TeV-PeV
neutrinos 
depend strongly on the spectrum index of the primary protons.
GeV-TeV gamma-rays will be detected by the 
ground-based Cherenkov telescopes such as CANGAROOIII and
the next-generation satellites such as GLAST when the spectrum of the
primary particles is soft $\gamma \sim 2.7$ and much of these cosmic rays are
mainly composed of neutrons as a result of the hadronic interactions.
Also, in such a case, TeV neutrinos will be detected at
the planned $1 {\rm km}^{2}$ detectors such as IceCube.
This means that detection of these signals from the Galactic Center
will constrain the spectrum of the primary particles, which, in turn,
should constrain the acceleration mechanism at the Galactic Center.
For example, if the spectrum is so flat and the intensity
of the GeV-TeV gamma-rays
is quite low, the simple Fermi acceleration mechanism will break down.
Such a flat spectrum will favor other acceleration mechanism
such as electro-magnetic acceleration inside the magnetosphere
around a neutron star~\cite{bednarek01}.


In this paper, most discussions are presented assuming 
that protons are accelerated at the Galactic Center, even though the
observed excess is not exactly the direction of the Galactic Center.
However, we only used the fact that the Galactic Center is surrounded
by a molecular gas whose mass is about $\sim 10^{8} {\rm M}_{\odot}$
to show that a significant amount of the cosmic rays should be
composed of neutrons due to the hadronic interactions.
That is why the results presented in this study should be
invalid even if the acceleration region is not the Galactic Center,
as long as the acceleration region is surrounded by a massive
molecular cloud~\cite{bednarek02}.
This problem on the location of the acceleration region will be
solved by the south AUGER which will start to be operated from 2004.


If the environment of the acceleration region is similar to that of AGN
or GRB, where charged particles are accelerated due to the Fermi mechanism
and the fields is filled with the synchrotron radiation, accelerated protons
will lose their energy due to the photopion interactions. Although the
fraction of energy lost by protons to pions is uncertain \cite{Waxman},
the resulting neutrino luminosity can be estimated as
\begin{equation}
L_{\nu} = 0.1 \left( \frac{f_{\pi}}{0.1} \right)
\frac{\frac{3}{13}}{1+\frac{3}{13}} L_{p} \;\;\; [\rm erg \; s^{-1}],
\label{estimate}
\end{equation}
where $f_{\pi}$ is the fraction of proton energy lost by photopion
interactions. Here we used that the ratio of neutrino to gamma luminosities
is approximately 1:3/13 when the $\Delta$ resonance occurs \cite{HALZEN}.
Since the cross section for the process $p \gamma \rightarrow p \pi^0$
and $p \gamma \rightarrow n \pi^{+}$ is about 2:1, the resulting ratio
of neutron to proton is
\begin{equation}
\frac{f_{n}}{f_{p}} 
 =  0.1 \times \left( \frac{f_{\pi}}{0.1} \right) \frac{1}{1+2} 
\sim 3.3 \times 10^{-2} \left( \frac{f_{\pi}}{0.1} \right).
\label{rat2}
\end{equation}
Since this ratio is about 10 times lower than the ratio assumed
in section~\ref{gamnudet}, the expected event rate will also
become about 10 times lower.

Neutrino event rate from the Galactic Center can be also estimated roughly
by comparing its activity to that of AGN. As the activity in
connection with neutrino, we can use the TeV-gamma-ray luminosity 
\cite{HALZEN}. Since there are no TeV-gamma-ray observation of
the Galactic Center, we compare GeV-gamma-ray luminosity.
This is justified because it is likely that the luminosity 
in TeV range is roughly the same 
as that in GeV range at the Galactic Center as is the case with TeV blazar 
\cite{Mrk421,Whipple}. We consider Mrk 421 as a representative example
for TeV blazar.
Using data from EGRET ( \cite{GC} for the Galactic
Center and \cite{Mrk421} for Mrk 421), the ratio of the neutrino
event rate for the Galactic Center ($N_{GC}$) to that for Mrk 421 ($N_{AGN}$)
is,
\begin{eqnarray}
\frac{N_{GC}}{N_{AGN}} & \sim & \frac{10^{36} {\rm erg/sec}}
                                  {10^{43} {\rm erg/sec}}
\frac{(10 {\rm kpc})^{-2}}{(100 {\rm Mpc})^{-2}} \\
                       & \sim & 10.
\end{eqnarray}
Since neutrino event rate for Mrk 421 is $\sim 0.03$ yr$^{-1}$
\cite{HALZEN,TA},
neutrino event rate for the Galactic Center is estimated to be
$\sim 0.3$ yr$^{-1}$.

When the neutrino mixing angles are really large~\cite{fukuda98}, the 
fluxes of $\nu_{\mu}$'s and $\bar{\nu}_{\mu}$'s will fall off
by a factor of 3. In this case, $\nu_{\tau}$'s might be detected
as the double bang events~\cite{athar00}.


It is very important to infer the location of the acceleration region
around the Galactic Center by $\gamma$-ray and/or X-ray observations,
which will give us an
information on candidates such as size, strength of magnetic fields
and electron density. 
The CANGAROO group is now surveying the Galactic Center to find
TeV gamma-ray source(s). The results will be soon reported 
(Kifune, private communication).
We will be able to judge the candidates using
these informations and Hillas diagram \cite{Hillas}.

In this study, we show that no bound is violated 
for gamma-ray flux in the range of GeV-TeV energies. However, there
may be even more severe bounds to this model from radio and X-ray
observations. In fact, the decay of charged pions also generate 
electron-positron pairs, radiating through the interstellar matter
and synchrotron emission in the buldge. To estimate the intensity
of these fluxes, Monte-Carlo simulations on the cascade showers of
high energy cosmic rays assuming some magnetic fields
will be needed. We are planning to perform such simulations and
the results will be presented in the near future.

In this study, we have estimated the fluxes of VHE gamma-rays and
neutrinos from the Galactic Center
which are by-products of the hadronic interactions
of primary protons which are accelerated up to $\sim 10^{19}$eV
and ambient matter. We found that these signals
depend strongly on the spectrum index of the primary protons.
We found that GeV-TeV gamma-rays will be detected by the ground-based
Cherenkov telescopes such as CANGAROOIII and the next-generation
satellites such as GLAST when the spectrum of the primary particles
is soft and much of the cosmic rays are mainly composed of neutrons.
In such a case, it is found that the TeV-PeV neutrinos will be also
detected by the planned $1 {\rm km}^{2}$ detectors such as IceCube.
Thus the acceleration region of the CRs below $\sim \; 10^{19}$eV will
be determined accurately by gamma-ray telescopes such as CANGAROOIII
and/or by neutrino detectors such as IceCube,
which should help our understanding on the particle acceleration.
Detection of these signals from the Galactic Center in the near future
will constrain the spectrum of the primary particles, which, in turn,
should constrain the acceleration mechanism at the Galactic Center.


\section{Acknowledgments}
The authors are grateful to Dr. M. Sasaki for useful discussions.
The authors are also grateful to the anonymous referee for his
useful comments.
This research has been supported in part by a Grant-in-Aid for the
Center-of-Excellence (COE) Research (07CE2002) and for the Scientific
Research Fund (199908802) of the Ministry of Education, Science, Sports and
Culture in Japan and by Japan Society for the Promotion of Science
Postdoctoral Fellowships for Research Abroad.

\newpage




\begin{table}
\caption{\label{tab:table0}
Event Rate (yr$^{-1}$) at IceCube within 1$^{\circ}$ circle}
\begin{tabular}{c|cccllcccccccc}
$\gamma$ & 2.7 & 2.0 & 1.5 & 1.0 \\
Event Rate (f=0.22)   & 6.5 $\times 10^{4}$ & 51 & 0.65 & 0.24\\
Event Rate (f=1.0)    & 3.0 $\times 10^{5}$ & 2.3$\times 10^{2}$  & 3.0 & 1.1\\
\end{tabular}
\end{table}

\end{document}